\documentclass{emulateapj}
\usepackage{epsfig}
\usepackage{natbib}
\usepackage{amsmath}
\usepackage{epstopdf}

\slugcomment{Published in ApJ Letters}
\shorttitle{Periodically Varying Quasar at $z = 2$ from PS1 MDS}
\shortauthors{Liu et al.}

\begin{document}

\title{A Periodically Varying Luminous Quasar at $\lowercase{z}=2$ from the Pan-STARRS1 Medium Deep Survey: A Candidate Supermassive Black Hole Binary in the Gravitational Wave-Driven Regime}

\author{Tingting Liu\altaffilmark{1},
Suvi Gezari\altaffilmark{1},
Sebastien Heinis\altaffilmark{1},
Eugene A. Magnier\altaffilmark{2},
William S. Burgett\altaffilmark{3},
Kenneth Chambers\altaffilmark{2},
Heather Flewelling\altaffilmark{2},
Mark Huber\altaffilmark{2},
Klaus W. Hodapp\altaffilmark{2},
Nicholas Kaiser\altaffilmark{2},
Rolf-Peter Kudritzki\altaffilmark{2},
John L. Tonry\altaffilmark{2}, 
Richard J. Wainscoat\altaffilmark{2},
and
Christopher Waters\altaffilmark{2}
}
\altaffiltext{1}{Department of Astronomy, 
        University of Maryland,
        College Park, MD 20742-2421, USA; \email{tingting@astro.umd.edu}}
        
\altaffiltext{2}{Institute for Astronomy, 
        University of Hawaii,
        2680 Woodlawn Drive, Honolulu, HI 96822, USA}
        
\altaffiltext{3}{GMTO Corporation, 251 S. Lake Avenue, Suite 300, Pasadena, CA 91101, USA}

\begin{abstract}

Supermassive black hole binaries (SMBHBs) should be an inevitable consequence of the hierarchical growth of massive galaxies through mergers and the strongest sirens of gravitational waves (GWs) in the cosmos. Yet, their direct detection has remained elusive due to the compact (sub-parsec) orbital separations of gravitationally bound SMBHBs. Here, we exploit a theoretically predicted signature of an SMBHB in the time domain:  periodic variability caused by a mass accretion rate that is modulated by the binary's orbital motion.  We report our first significant periodically varying quasar detection from the systematic search in the Pan-STARRS1 (PS1) Medium Deep Survey.  Our SMBHB candidate, PSO J334.2028+01.4075, is a luminous radio-loud quasar at $z=2.060$, with extended baseline photometry from the Catalina Real-Time Transient Survey, as well as archival spectroscopy from the FIRST Bright Quasar Survey.  The observed period ($542 \pm 15$ days) and estimated black hole mass ($\log (M_{\rm BH}/M_\odot) = 9.97 \pm 0.50$), correspond to an orbital separation of $7^{+8}_{-4}$ Schwarzschild radii ($\sim 0.006^{+0.007}_{-0.003}$ pc), assuming the rest-frame period of the quasar variability traces the orbital period of the binary.  This SMBHB candidate, discovered at the peak redshift for SMBH mergers, is in a physically stable configuration for a circumbinary accretion disk and within the regime of GW-driven orbital decay.  Our search with PS1 is a benchmark study for the exciting capabilities of LSST, which will have orders of magnitude larger survey power and will potentially pinpoint the locations of thousands of SMBHBs in the variable night sky.\\
\end{abstract}

\keywords{quasars: general --- surveys }

\section{Introduction}  \label{intro}

The expectation for the existence of supermassive black hole binaries (SMBHBs) in galaxy nuclei is supported by two well-established properties of galaxies:  (1) high spatial resolution observations of nearby galaxies have demonstrated that SMBHs are ubiquitous in the centers of galaxy bulges \citep{Magorrian1998} with masses tightly correlated with the mass and structure of their host galaxies \citep{Ferrarese2000, Gebhardt2000, Graham2001}, and (2) galaxies in a $\Lambda$CDM Universe build up their structure hierarchically through mergers \citep[e.g.][]{Springel2005}.  When two galaxies merge, their SMBHs will sink to the center through dynamical friction and through three-body interactions with stars and viscous exchange of angular momentum with circumbinary gas form a gravitationally bound binary that eventually coalesces due to the radiation of gravitational waves (GWs; \citealt{Begelman1980}).  

Recent progress has been made in the detection of ``dual active galactic nuclei (AGNs)," double active nuclei in assumed merged galaxy systems with kiloparsec-scale separations \citep{Komossa2003, Comerford2009}.  These dual AGNs, while a product of a galaxy merger, are not yet gravitationally bound, and thus are not necessarily fated to coalesce.   A true SMBHB becomes gravitationally bound on the scale of parsecs, which, beyond our Local Group of galaxies, is well below the angular resolving power of the most powerful current, or even future, telescopes.  However, several promising candidates have been identified indirectly via spectroscopy: quasars with offset and/or drifting broad-line peaks attributed to a broad-line region in orbit around an SMBH's binary companion \citep{Boroson2009, Dotti2009, Barrows2011}.  However, alternative scenarios have been proposed that do not require an SMBHB, including double-peaked lines from a single accretion disk \citep{Chornock2010}.  

A promising observational signature of SMBHBs is their variable accretion luminosity. One of the first sub-parsec SMBHB candidates, OJ287, was identified by its variability behavior \citep{Lehto1996}.  OJ287 is a quasar that demonstrates regular optical outbursts on a timescale of 12 yr that has been modeled as the result of a secondary SMBH companion passing through the primary SMBH's accretion disk \citep{Valtonen2008}.  Such a configuration should be rare since the secondary BH's orbital axis must be highly misaligned with the primary BH's accretion disk axis in order for it to pass through its disk.    A more generic signature of an SMBHB is likely to be related to accretion through its circumbinary disk. 

In a gas-rich galaxy merger, strong gravitational torques drive gas inward, triggering both star formation and BH accretion \citep{Hopkins2006}.  In particular, hydrodynamical simulations of circumbinary disks show that accretion via ``hot streams'' onto the BHs is strongly modulated by the binary's orbital motion for mass ratios of $0.05<q\le1$ \citep{MacFadyen2008, Shi2012, Noble2012, DOrazio2013, Gold2014}.  Simulations \citep{DOrazio2013, farris2015a} also detect a $t \sim 6$ $t_{\rm orb}$ timescale originating from a surface density ``lump'' just outside the central cavity of the circumbinary disk, which is a persistent but secularly evolving feature. 

While working on our paper, we became aware of the report of PG 1302-102, a periodically variable quasar discovered by the Catalina Real-Time Transient Survey (CRTS; \citealt{graham2015}). It is a $15$ magnitude quasar at $z = 0.2784$, varying at the 0.14 mag level with a period of $5.2\pm0.2$ yr, with good sampling over 1.8 cycles and extended archival data going back 20 yr. Their physical interpretation for its variability is an SMBHB ($\log (M/M_\odot) \sim 8.5$, $a \sim 0.01$ pc), its luminosity being modulated due to either a precessing jet or an overdensity (``hot spot'') in the inner edge of its circumbinary disk.

In this Letter, we present our most significant detection from the systematic search for periodically varying quasars in the Pan-STARRS1 Medium Deep Survey (PS1 MDS) field MD09.   PSO J334.2028+01.4075 is a radio-loud quasar at $z=2.060$ with archival spectroscopy from FQBS and extended baseline photometry from CRTS.  The 8.5 yr baseline of the PS1+CRTS light curve is well described by a simple sinusoid, consistent with theoretical simulations for the modulated accretion rate in a $0.05 < q < 0.25$ mass-ratio SMBHB.  We use the rest-frame period and virial black hole mass estimate to infer an orbital separation of the binary that is in the GW-driven regime.

\section{Theoretical Predictions} \label{sec_theory}

The dynamics of a gravitationally bound SMBHB system can be described by Kepler's third law:

\begin{equation}
	\label{eqn:timescale}
	t_{\rm orb} = 0.88\,\mbox{yr} \left(\frac{M}{10^{7} M_\odot}\right) \left(\frac{a}{10^3 R_{s}}\right)^{3/2}\quad, 
\end{equation}

\noindent where $R_s$ is the Schwarzschild radius, $R_{\rm s} = \frac{2GM}{c^2}$; $a$ is the separation between the BHs; and $M$ is the total mass of the system.  \cite{Haiman2009} calculate the minimum survey area to detect a statistically significant sample of quasars powered by SMBHBs as a function of variable magnitude depth, assuming reasonable values for the quasar luminosity function, quasar lifetime ($t_Q = 10^7$ yr), the Eddington fraction ($f_{\rm Edd} = 0.3$), and the fractional variability amplitude ($\Delta f/f = 0.1$).  The variability detection threshold we have achieved in MD09 (\S \ref{sec_ensemble}) corresponds to a $\Delta f/f > 0.1$ sensitivity for point sources brighter than $m \sim 21$ mag, and thus a variable magnitude of 23.5 mag. At this depth, \cite{Haiman2009} require an area of $\sim $100 deg$^2$ to yield a sample of over 100 SMBHBs; an excellent match to the area of PS1 MDS (80 deg$^2$).  Furthermore, the baseline (4.2 yr) and cadence (3 days) of PS1 MDS make us sensitive to timescales for which BHs (with $M > 10^7 M_\odot$) are in the GW-driven regime of orbital decay.

\section{Pan-STARRS1 Medium Deep Survey}\label{sec_ps1}

The Panoramic Survey Telescope and Rapid Response System (Pan-STARRS) is a wide-field imaging system designed for dedicated survey observations on a 1.8m telescope on Haleakala, Hawaii, with a 1.4 Gigapixel camera and
a 7 deg$^{2}$ field of view \citep{Kaiser2010}.  The PS1 telescope is operated by the Institute for Astronomy (IfA) at the University of Hawaii and has just completed over $4$ yr of operation in 2014 March.  We present data from the Medium Deep Survey (MDS), a deep, multi-epoch survey of $10$ circular fields distributed across the sky, each $\sim 8$ deg$^{2}$ in size, whose daily observing cadence in five filters is excellent for studying persistent variable sources, including quasars.
The PS1 MDS cadence of observation cycles through the $g_{\rm P1}$, $r_{\rm P1}$, $i_{\rm P1}$, and $z_{\rm P1}$\ bands every three nights, with observations in the $y_{\rm P1}$\ band close to the full Moon.  Due to the poorer time sampling of the $y_{\rm P1}$ observations, we do not use them in this analysis.

\section{Ensemble Photometry}\label{sec_ensemble}

\begin{figure*}[h]
	\centering
	\epsfig{file=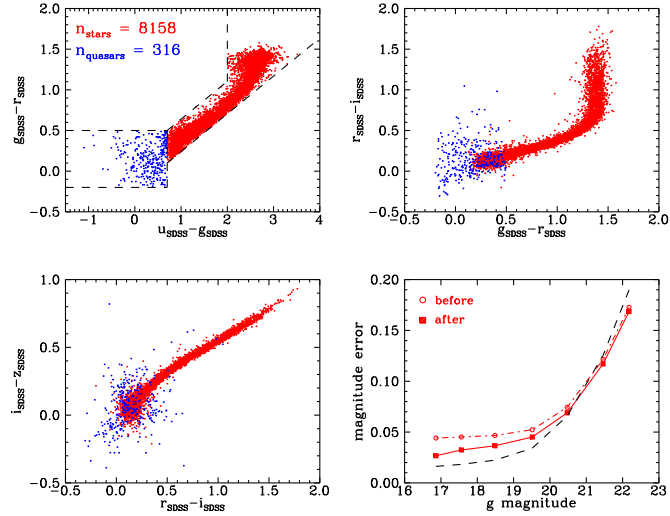,width=0.5\textwidth,clip=}
	\caption{Color--color diagrams used to select point sources in MD09 for the 8158 point sources in the reference star sample (red) and 316 point sources in the quasar sample (blue).  Photometry is measured from the CFHT+PS1 catalog and converted to the SDSS system.  Dashed lines show the color selection boundaries.  The stellar color-color selection box was chosen to avoid RR Lyrae stars, which are intrinsically variable.  Bottom right panel: observed standard deviation of the reference sample of non-variable stars before and after applying the technique of ensemble photometry (open circles and filled squares, respectively), compared to the Poisson error expected from the reported PS1 flux errors (black dashed lines). }
	\label{fig:colors}
\end{figure*}

We began our systematic search for SMBHB candidates among color-selected quasars in the PS1 MD09 field. This is the first MD field that was made available to the PS1 Science Collaboration in the Pan-STARRS Science Interface (PSI) online database.  In order to maximize our sensitivity to intrinsic variability, we first applied the technique of differential ensemble photometry.  This technique is able to correct for local systematic errors due to variable atmospheric conditions by comparing a target object with nearby non-variable stars \citep{honeycutt1992, Bhatti2010}.  We created a color-selected reference star sample and quasar sample by cross-matching point sources ($m < 23$ mag) in the MD09 field with a custom catalog extracted from full-survey deep stacks from PS1 MDS in the $g_{\rm P1}$, $r_{\rm P1}$, $i_{\rm P1}$, $z_{\rm P1}$, and $y_{\rm P1}$\ bands, as well as from observations with the Canada--France--Hawaii Telescope (CFHT) in the $u$ band (S. Heinis et al. 2015, in preparation).   We converted all magnitudes to the SDSS photometric system \citep{tonry2012} to take advantage of the SDSS color selection of stars and quasars already available in the literature \citep{Schmidt2010, Sesar2007}.  Figure \ref{fig:colors} shows the color-color diagrams of the point sources in MD09 selected as quasars and non-variable stars.  This query resulted in 8158 reference stars and 316 quasars, each with an average of 350 detections in four filters.

We modified the ensemble photometry software developed by \citet{Bhatti2010} for SDSS to the PS1 data format and ran it on the reference stars.  In Figure \ref{fig:colors} (bottom right panel), we plot the ``corrected'' magnitude error as a function of mean magnitude compared with the ``raw'' values before ensemble photometry. The ensemble photometry reduces the measured errors significantly, lowering the error floor from 0.045 to 0.025 mag, and resulting in a 2$\sigma$ variability threshold of $0.05$ mag on the bright end to 0.34 mag on the faint end.

\section{Selecting Periodic Quasar Candidates}\label{sec_cand}

We then applied ensemble photometry to the 316 color-selected quasars and flagged quasars as variable based on their magnitude error relative to their neighbors of a similar brightness; we set $2 \sigma$ as our criterion for variability and required a variability flag in at least two filters. This selection yielded $168$ variable quasars in MD09.
Among these color-selected variable quasars, we searched for potential periodic signatures using the Lomb--Scargle (LS) periodogram, a Fourier analysis technique of unevenly spaced data with noise \citep{Lomb1976, Scargle1982, Horne1986}. For $N_0$ data points in the time series spanning a total length of $T$ in units of MJD, we sampled the periodogram at the number of recommended independent frequencies ($N_i$), from \cite{Horne1986}, from $1/T$ to $N_0/(2T)$ (which would be the Nyquist frequency if data were evenly sampled), resulting in a frequency resolution in the periodogram of $\Delta f = (N_0/2-1)/(TN_i) \sim 2\times10^{-4}$ d$^{-1}$.   When identifying periodic sources, we took advantage of the redundancy of PS1 MDS monitoring in four filters ($g_{\rm P1}$\, $r_{\rm P1}$\, $i_{\rm P1}$\, $z_{\rm P1}$), each with a slightly different observing cadence due to weather and scheduling constraints to help filter out false detections from aliasing by requiring that periodogram peaks be coherent across multiple filters; $40$ of the $168$ variable quasars survived this test. 

\section{Periodic Quasar Candidate PSO J334.2028\\+01.4075}\label{sec_archival}

\begin{figure*}[h]
	\centering
	\epsfig{file=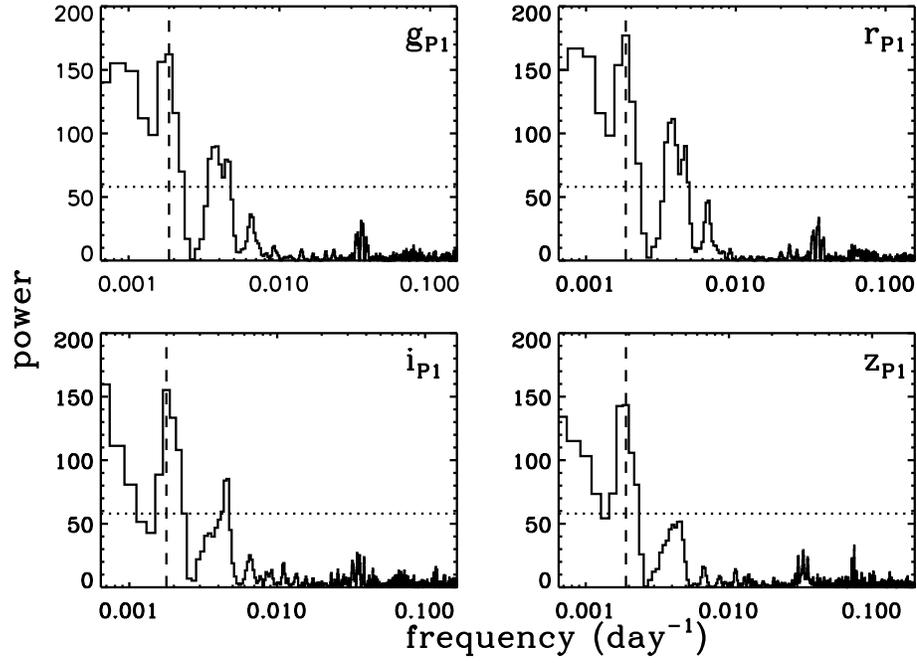,width=0.75\textwidth,clip=}
	\caption{Our automated LS periodogram routine selects periodically variable candidates by requiring that the strongest peak is detected at the same frequency in at least three filters. This quasar candidate, PSO J334.2028+01.4075, was selected through this method and had the periodogram peak with the highest signal-to-noise ratio of all of our candidates.  The dashed lines mark the strongest peak in each filter. The dotted line corresponds to a false-alarm probability of $1.5\times10^{-23}$, or 10 $\sigma$.}
	\label{fig:ps1_pgrams}
\end{figure*}

\begin{figure*}[h]
	\centering
	\epsfig{file=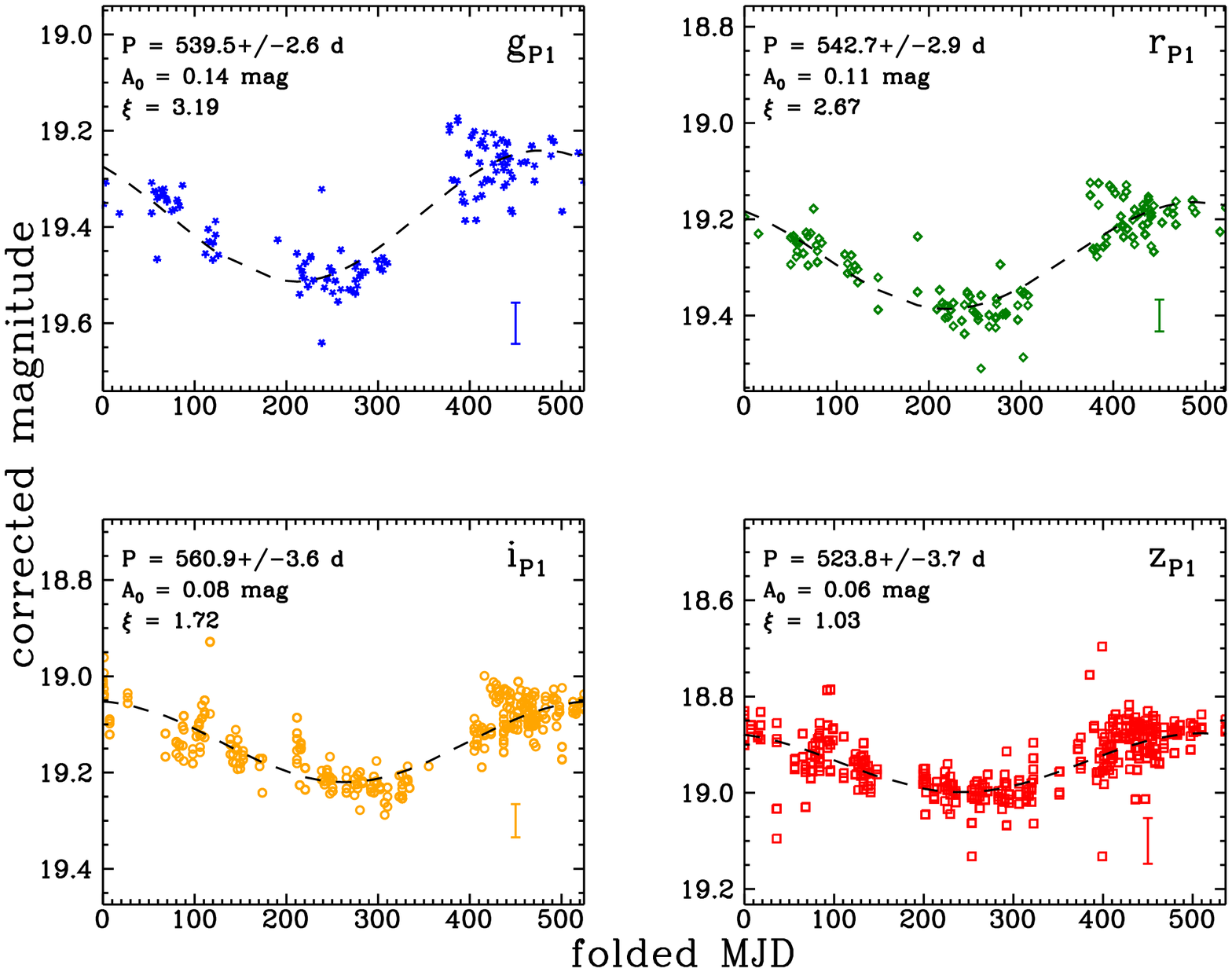,width=0.7\textwidth,clip=}
	\epsfig{file=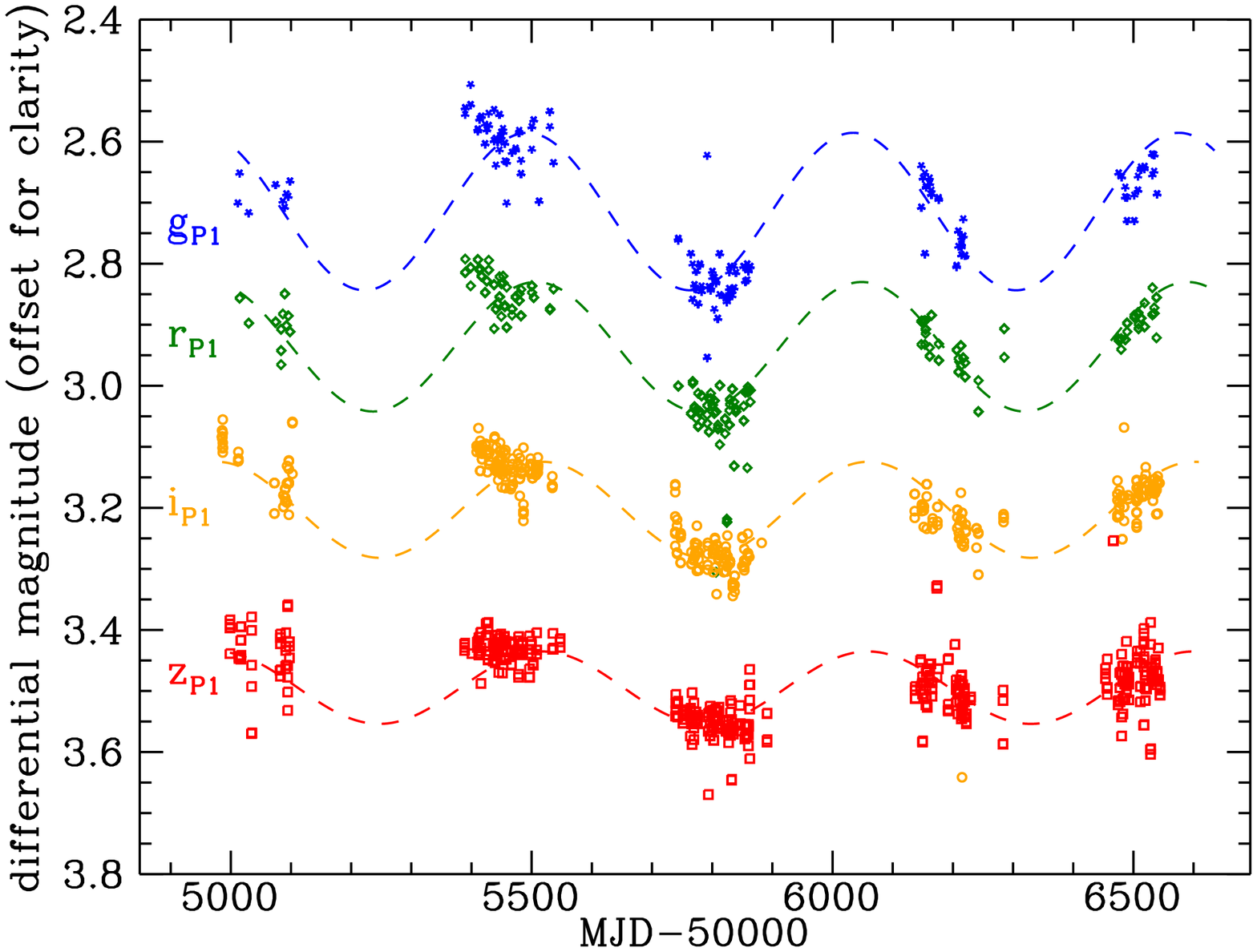,width=0.77\textwidth,clip=}
	\caption{Upper panels:  sinusoidal fit to the folded PS1 light curve of PSO J334.2028+01.4075 in four filters, with the error bar indicating the typical photometric error for an object of similar brightness in that filter.  The period corresponding to the peak of the periodogram and its error bar, the amplitude of the fitted sine wave, and the signal-to-noise ratio of the peak power are each labeled.  Lower panel: sinusoidal fit plotted over the complete PS1 light curves in the $g_{\rm P1}$\, $r_{\rm P1}$\, $i_{\rm P1}$\, and $z_{\rm P1}$ bands. The data used to create this figure are available.}
	\label{fig:fit_3783}
\end{figure*}

Among the candidate periodic quasars from our periodogram analysis, here we focus on our most significant detection, PSO J334.2028+01.4075 (J2000).  In Figure \ref{fig:ps1_pgrams}, we show its periodogram in four PS1 filters, with the strongest peak marked with a dashed line.  We fold each filter light curve on this period (Figure \ref{fig:fit_3783}), and measure the scatter of the residuals from the best-fit sine curve ($\sigma_r$).   The error of the periodogram peak frequency ($\delta f$) and the signal-to-noise ratio of the peak power ($\xi$) can then be calculated from $\sigma_r$ and the amplitude of the signal $A_0$ \citep{Horne1986} as $\delta f = \frac{3\sigma_r}{4\sqrt{N_0}TA_0}$ (which gives us an error on the detected period of $\delta P = \delta f/f^2$) and $\xi \equiv A_0^2/(2 \sigma_r^2)$, respectively.  

The resulting average period across all four filters is $P = 541.8 \pm 15.3$ days, with the highest signal-to-noise ratio in the $g_{\rm P1}$\ filter with $\xi = 3.19$, and periodogram peaks in all four filters well above a $1.5\times10^{-23}$ false-alarm probability (corresponding to 10$\sigma$) plotted with a dotted line in Figure \ref{fig:ps1_pgrams}.  The PS1 data cover 2.6 cycles, just shy of the ``rule of thumb" number of cycles (three) for a periodic variation to be apparent to the eye \citep{press1978}. From our Monte Carlo simulations of $1000$ stochastic Damped Random Walk \citep{Kelly2009} light curves, we find a false periodic detection rate of $6.3 \%$ using our selection criteria from \S 5. We further disfavor a false alarm from stochastic quasar variability since the $0.6 \%$ of the simulations that successfully mimic the periodic timescale of our candidate have short-timescale variances a factor of $\gtrsim 2$ larger than expected for the quasar's luminosity and inferred black hole mass (\S 7). Note that there is a secondary peak in the $g_{\rm P1}$ and $r_{\rm P1}$ periodograms that, if real and not an artifact from the PS1 data sampling, is at twice the primary peak frequency, a signature of $0.05 < q < 0.25$ mass-ratio SMBHBs, which show an accretion rate modulation most closely described by a simple sinusoid \citep{DOrazio2013}. The amplitude of PSO J334.2028+01.4075's sinusoidal modulation increases with decreasing wavelength, consistent with the exponential dependence on wavelength found in previous quasar variability studies (e.g.\ \citealt{vandenberk2004, macleod2010}).  

\begin{figure*}[h]
	\centering
	\epsfig{file=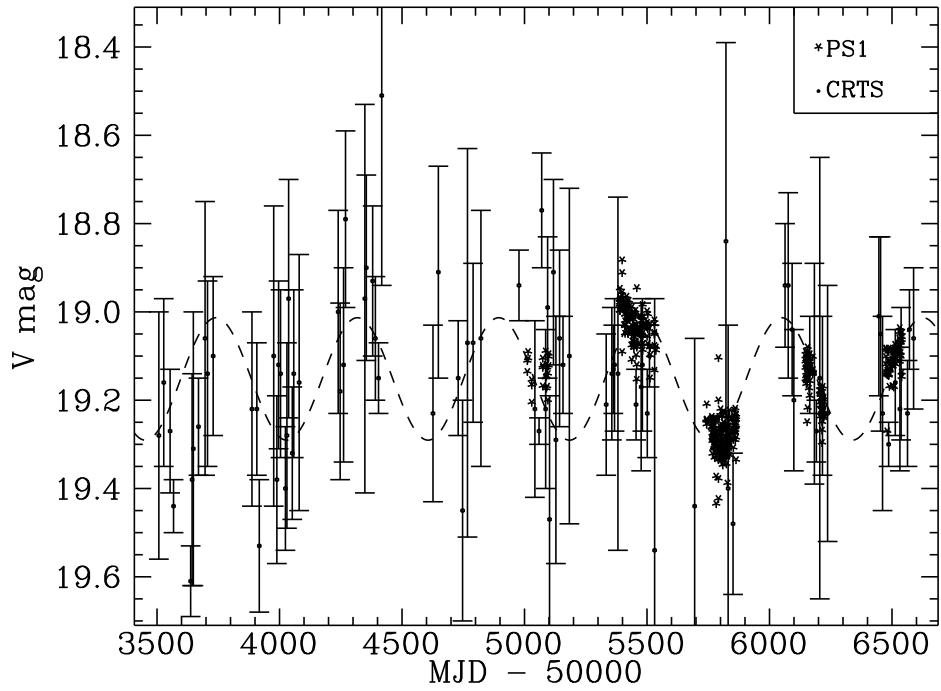,width=0.75\textwidth,clip=}
	\caption{PS1 light curve (asterisks) converted to the $V$ band to compare to the archival CRTS light curve (dots with error bars).  The CRTS data points are binned in one-day intervals, with the error bars measured from the standard deviation in the bin and not including data points with a photometric error greater than 0.25 mag or nights with less than three measurements. This results in 34/113 nights of data being thrown out. The CRTS measurements are overall consistent with the PS1 light curve and the sine fit to the PS1 light curve (dashed curve).}
	\label{fig:crts_3783}
\end{figure*}

PSO J334.2028+01.4075 is a radio-loud quasar (FBQS J221648.7+012427) with an archival spectrum from the FIRST Bright Quasar Survey \citep{Becker2001}.  We are also fortunate that this candidate has an archival V-band light curve from CRTS \citep{drake2009}, which we use to test the persistence of the periodic variation over an extended baseline of 8.5 yr (corresponding to 5.7 cycles). To compare to the CRTS light curve, we convert the PS1 $g_{\rm P1}$-band light curves to the SDSS system \citep{tonry2012} and then to the Johnson $V$ magnitude using the photometric transformation for quasars from \cite{Jester2005} and an average $g_{\rm P1}$-$r_{\rm P1} = +0.10$ mag.  We had to apply an additional offset of $-0.17$ mag to the pseudo-$V$ PS1 magnitudes in order to match the average of the CRTS $V$-band data.   Though the photometric errors are relatively large, the CRTS measurements are consistent with those of PS1 during their overlap (Figure \ref{fig:crts_3783}), and have residuals over the entire CRTS baseline from the PS1-fitted sinusoidal model that are Gaussian with a $\sigma = 0.17$ mag that is comparable to the mean photometric error of 0.18 mag. 

\section{Physical Interpretation}\label{sec_model}

We use the width of the quasar's $C$ IV line and its nearby continuum luminosity to make a virial estimate of the black hole mass from \citet{vp2006}:

\begin{equation}
	\log\Big(\frac{M_{\rm BH}}{M_\odot}\Big)  = \log\Big[\Big(\frac{\text{FWHM(C IV)}}{1000 \text{km/s}}\Big)^2 \Big(\frac{\lambda L_\lambda}{10^{44} \text{erg/s}}\Big)^{0.53} \Big] + 6.66
	\label{eqn:vestergaard}
\end{equation}

\noindent where $\lambda = 1350$ \AA\ .  The FBQS spectrum, though not publicly available in an electronic format, was measured with a ruler to determine $F_{\lambda,{\rm obs}}$(1350\AA\ (1+z)) $\sim 8.5 \times 10^{-17}$ ergs s$^{-1}$cm$^{-2}$ \AA$^{-1}$, and FWHM (CIV $\lambda 1550$) $\sim 200$ \AA.  The CIV line is symmetric in shape, and its width corresponds to a velocity in the rest-frame of $\sim 12,650$ km s$^{-1}$.  We correct for a Galactic extinction of $E(B-V) = 0.0406$ mag \citep{sf2011}, using the extinction law from \cite{cardelli1989}, to find $L_{\lambda, \rm em}(1350 \AA) = 4 \pi d_L^2 F_{\lambda, \rm obs} 10^{A_\lambda/2.5} (1+z) \sim 9.5 \times 10^{42}$ erg s$^{-1}$\AA$^{-1}$, where $d_L$ is the luminosity distance assuming $H_0 = 70$ km/s/Mpc, $\Omega_m = 0.3$, $\Omega_\Lambda = 0.7$, and $k = 0$, and $A_\lambda = 0.1790$.  This results in a black hole mass of $\log\Big(\frac{M_{\rm BH}}{M_\odot}\Big) = 9.97$, with a scatter from the uncertainty in the relation of 0.5dex.
Applying a mean quasar bolometric correction at $1350$\AA\ of BC=$3.81$ from \citet{Richards2006}, one gets a bolometric luminosity of $L_{\rm bol} = \lambda L_\lambda BC = 4.9\times 10^{46}$ erg s$^{-1}$.  
Note that this object is also radio loud, with a radio luminosity at the rest-frame frequency of 5 GHz of $\log(L_R $(erg s$^{-1}$)) = 32.8 from \cite{Becker2001}.  

Assuming the rest-frame period $P_{\rm rest} = P_{\rm obs}/(1+z)$ is on the order of the orbital timescale of the SMBHB, with a caveat that in addition to a strong dependence on mass ratio, there are a range of theoretical predictions for translating $P_{\rm rest}$ to $t_{\rm orb}$ (e.g.\ \citealt{Noble2012}, \citealt{DOrazio2013}), we then calculate the orbital separation of the binary to be $7^{+8}_{-4} R_s$ ($\sim 0.006^{+0.007}_{-0.003}$ pc), securely placing it in the gravitationally bound regime of a physically viable SMBHB system --- a circumbinary accretion disk system capable of maintaining a central cavity, stable to gravitational fragmentation, and in the regime of orbital decay driven by GWs (\citealt{Haiman2009, kocsis2012, DOrazio2013}).  Also note that since the viscous time scales as $r^2$, one could expect to be able to detect modulations in the accretion rate fed by the streams in the circumbinary disk cavity, without being washed out by viscous processes in the ``mini disks'' around each BH \citep{Roedig2014}.  Remarkably, the rest-frame inspiral time for the binary is $t_{\rm insp} = \frac{5}{256}\frac{c^5}{G^3}\frac{a^4}{M^2\mu} = 7.0\, \rm yr\,(a/7 R_s)^4 (M/10^{9.97} M_\odot)^{-3}$ for $q = 0.25$, where $\mu \equiv \frac{M_1 M_2}{M_1 + M_2}$, opening up the possibility for detecting the decay of the orbital period $(\dot P)$ with future monitoring, as well as providing a promising target for direct GW detection for pulsar timing arrays \citep{Sesana2009}.

\section{Discussion and conclusions}\label{sec_discussion}

We present the most statistically significant periodically variable quasar candidate from our search in PS1 MD09, PSO J334.2028+01.4075, a radio-loud quasar at $z=2.060$.   We combine an estimate of its black hole mass with its variability timescale (assuming $P_{\rm rest} \sim t_{\rm orb}$) to find orbital parameters consistent with model predictions of a stable accreting SMBHB system with a $0.05 < q < 0.25$ in the GW-driven regime (\citealt{Haiman2009}).  

The redshift of this SMBHB candidate coincides with the peak epoch for SMBHB mergers \citep{Volonteri2003}, and its large mass ($M \approx 10^{10} M_{\odot}$) makes it favorable for detection in the GW-driven regime, due to the strong dependence on $M$ of the residence time at a given orbital separation in units of $R_s$ \citep{Haiman2009}.  Like the CRTS SMBHB candidate PG 1302-102 reported by \citet{graham2015}, our SMBHB candidate is also a radio-loud quasar.  However, given the shorter rest-frame period of our candidate of 0.5 yr (versus 4 yr in PG 1302-102), it is even more unlikely for its variability to be driven by jet precession, either originating from a single SMBH \citep{Lu2005} or a binary SMBH \citep{Lobanov2005}.

This pilot program in PS1 MD09 is a promising start to our systematic search for periodic variability signatures of SMBHBs amongst the expected $\approx 1000$ variable quasars in the full $\sim 80$ deg$^{2}$ PS1 MDS.  At the start of the next decade ($\sim$ 2023), the Large Synoptic Survey Telescope (LSST) \citep{Ivezic2008} will probe a volume several thousand times larger than PS1 MDS, yielding tens of millions of quasars, and potentially thousands of SMBHBs periodically varying on the timescale of years, fated to coalesce.  

\acknowledgements
S.G. thanks the Aspen Center for Physics, which is supported in part
by the NSF under grant No. PHYS-1066293, for hosting the ``Black Holes in Dense Star Clusters'' Winter Workshop which facilitated stimulating discussion of SMBHB theory.  T.L. thanks J.~Krolik, R.~Mushotzky, and the anonymous referee for helpful comments on the manuscript.

The Pan-STARRS1 Surveys (PS1) have been made possible through contributions of the IfA, the University of Hawaii, the Pan-STARRS Project Office, the Max Planck Society and its participating institutes, MPIA, Heidelberg and MPE, Garching, JHU, Durham University, the University of Edinburgh, QUB, the Harvard-Smithsonian CfA, LCOGT Inc., the National Central University of Taiwan, STScI, NASA under grant No. NNX08AR22G issued through the Planetary Science Division of the NASA Science Mission Directorate, NSF under grant No. AST-1238877, the University of Maryland, and Eotvos Lorand University.

%\bibliographystyle{fapj}
%\bibliography{ms.bib}

\end{document}